# Modulated Hawking radiation and a nonviolent channel for information release

Steven B. Giddings*

*Department of Physics*

*University of California*

*Santa Barbara, CA 93106*

**Abstract**

Unitarization of black hole evaporation requires that quantum information escapes a black hole; an important question is to identify the mechanism or channel by which it does so. Accurate counting of black hole states via the Bekenstein-Hawking entropy would indicate this information should be encoded in radiation with average energy flux matching Hawking's. Information can be encoded with no change in net flux via fine-grained modulation of the Hawking radiation. In an approximate effective field theory description, couplings to the stress tensor of the black hole atmosphere that depend on the internal state of the black hole are a promising alternative for inducing such modulation. These can be picturesquely thought of as due to state-dependent metric fluctuations in the vicinity of the horizon. Such couplings offer the prospect of emitting information without extra energy flux, and can be shown to do so at linear order in the couplings, with motivation given for possible extension of this result to higher orders. The potential advantages of such couplings to the stress tensor thus extend beyond their universality, which is helpful in addressing constraints from black hole mining.

---

* Email address: giddings@physics.ucsb.edu

# 1. Introduction

If quantum mechanics governs nature, formation and decay of a black hole (BH) must be a unitary process. Local quantum field theory (LQFT) evolution on the semiclassical BH background[1] in contrast predicts dramatic loss of information. Unitarization of this evolution apparently requires significant new physics beyond such a LQFT description.

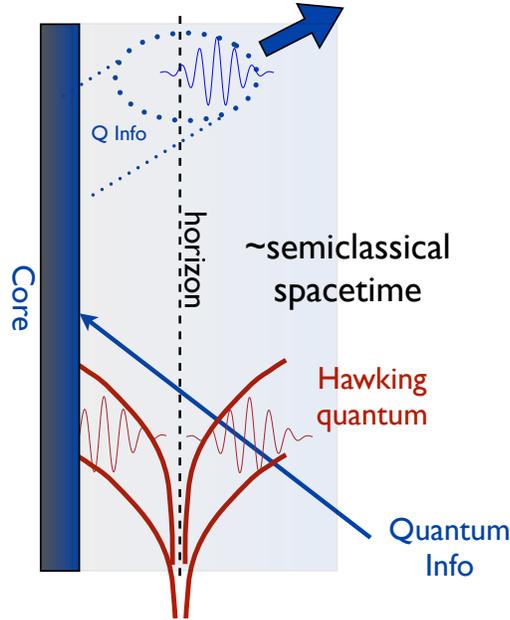

**Fig. 1:** Proposed schematic picture of unitary black hole evolution. Entanglement is built up between the BH and exterior through infall of quantum information or the Hawking process. Semiclassical evolution fails at the "core" of the BH. New effects extending through a region (shaded) including the BH atmosphere transfer information into outgoing modes that then escape, while preserving, to a good approximation, the semiclassical BH geometry.

Key questions are where and how such modifications of this LQFT evolution become relevant. In particular, we expect that LQFT in semiclassical spacetime geometry furnishes a good approximate description of physics far from a BH. On the other hand, LQFT is expected to be strongly corrected in the deep interior, or core, of BH. This by itself appears insufficient to transfer information out of the BH and unitarize evolution. But, if there are also small corrections to LQFT in an intermediate region – the immediate vicinity or atmosphere of the BH – extending outside the horizon (see Fig. 1), these offer the prospect of unitarizing evolution.

An illuminating way to describe information transfer is via transfer of entanglement[2-4]. A BH builds up entanglement with its environment either by absorbing matter entangled with the surroundings, or through production of Hawking particles entangled with interior excitations. Unitarity requires that all this entanglement ultimately transfers out, so that the fine-grained von Neumann entropy $S_{\rm vN}$ of pre- and post-BH states are equal. A simplest example of such transfer is just transfer of degrees of freedom[3]. A critical question, then, is what mechanism or dynamics is responsible for this transfer; such a mechanism appears beyond usual LQFT dynamics. In information-theoretic terms, we can frame the issue by focussing on the question: *what channel is responsible for escape of the information from the BH interior to infinity?* This approach contrasts with, *e.g.* [5-8], where LQFT is instead altered by modifying the property of *localization* of information.

There are many constraints on possible channels. In particular, if LQFT is exactly valid outside the horizon, such information transfer produces singular behavior at the horizon[9-12]. For this reason, it appears important that corrections to LQFT reach beyond the horizon, and the proposal that such corrections yield "nonviolent" transfer of information, preserving usual spacetime near the horizon to a good approximation, has been made and investigated in [13-17].

Even with such exterior corrections to LQFT, the general intuition that information transfer requires energy transfer appears borne out, and generic models for information escape produce extra energy flux beyond Hawking's[13-18]. This would indicate[19] that the internal states of a BH are not accurately parameterized by the Bekenstein-Hawking entropy $S_{\rm BH}$. While this may be consistent[19], a more conservative and appealing alternative would be no extra flux.

In an asymptotic LQFT description, we certainly expect that there are quantum states with coarse-grained thermal properties of the Hawking radiation, but with vanishing $S_{\rm vN}$. This motivates the proposal that the radiation carries the information in fine-grained modulations, with average energy flux matching Hawking's. An important question is whether this flux arises from perturbations of the Hawking radiation. A possible loose analogy is that of modulation of a radio signal: an underlying carrier flux can be modulated to transmit information.

Specifically, one can investigate whether such modulation of the Hawking radiation can be induced, while avoiding destruction of the horizon[9-12]. An important question is what channel or mechanism imprints the information on the outgoing radiation yet preserves near-horizon spacetime. This paper will propose and investigate a candidate mechanism/channel for information to flow from the BH interior to asymptotic observer, which offers the possibility of avoiding extra net flux.

In particular, if such dynamics can be approximately parameterized as a small correction to LQFT, it might be described by couplings to near-horizon fields that depend on the state of the BH[15,16]. It is desirable for such couplings to be universal, in particular to address mining constraints[12,15], and this suggests coupling the internal state to external fields through their stress tensor[15]. Interestingly, we will find that these couplings avoid producing leading-order corrections to energy flux. Such couplings through the stress tensor can be picturesquely thought of as due to near-horizon metric fluctuations correlated with the internal state of the BH. While the *fundamental* picture is not expected to be via such nonlocal corrections to LQFT, this may be for present purposes a good *approximate* description of a more fundamental dynamics[14]. This paper will explore such a description. A more complete description is possibly based on a fundamental tensor-factor structure[20,14].

## 2. States and evolution: scrambling and transfer

Assume that a BH coupled to its environment is represented in terms of states in a product space, with factors corresponding to BH and environment subsystems; this is a coarsest decomposition of the overall Hilbert space, and more refined versions may be considered[14,19]. Let $\widehat{|I\rangle}$ denote a basis of internal states $\mathcal{H}_M$ for a BH of mass $\leq M$, and suppose there are $\exp\{S_{\text{bh}}(M)\}$ such states. The entanglement entropy $S_{\text{vN}}$ of BH with environment is bounded above by $S_{\text{bh}}(M)$. According to LQFT, the entanglement $S_{\text{vN}}$ increases continually in the Hawking process[1]. But, once $S_{\text{vN}}$ reaches $S_{\text{bh}}$, decrease of $S_{\text{bh}}$ with $M$ means that transfer of entanglement from BH interior to exterior must take place to preserve unitary evolution. Thus, one postulates[21-23,13-15] couplings that transfer information from the BH interior to environment. Within LQFT these would be forbidden by prohibition of superluminal signaling with respect to the semiclassical geometry.

The presence of such effects may provide a critical clue to the underlying nature of quantum gravity. For now we give a general approximate parameterization[15-17] of them in terms of couplings between the BH states and the states $\mathcal{H}_{\text{near}}$ in the near-BH atmosphere. The latter states are expected to be approximately described within LQFT. Consider a hamiltonian description. LQFT evolution in the Schrödinger picture produces pairs of Hawking excitations in $\mathcal{H}_M \otimes \mathcal{H}_{\text{near}}$, transfers excitations from $\mathcal{H}_{\text{near}}$ to $\mathcal{H}_M$ (infall), and describes interactions between $\mathcal{H}_{\text{near}}$ and far states in $\mathcal{H}_{\text{far}}$.

The non-LQFT completion of this evolution necessary to restore unitarity may involve two other processes[14]. The first is *scrambling*, which can be described in terms of unitary evolution mixing internal states,

$$\widehat{|I\rangle} \to U_{IJ}(t)\widehat{|J\rangle} \ . \qquad (2.1)$$

$U_{IJ}$ is expected to depend on gauge[15]; for example, in LQFT evolution, in a gauge corresponding to a nice slicing[24,14], evolution of internal states freezes[25], implying $U = 1$ in this approximation.[1] In fact, while the internal Hilbert space and evolution $U$ are sometimes modeled as generic[26], we expect that they have special properties, since the evolution of internal states should describe observations of infalling observers in what to them initially appears to be weakly-curved space, so should approximately match such a LQFT description for those observers.

The second new process is *transfer* of information from $\mathcal{H}_M$ to $\mathcal{H}_{\text{near}}$. It can be written in terms of couplings in the action or hamiltonian (the latter being the generator of unitary evolution, in some slicing or gauge) between operators $\mathcal{A}_a$ acting on $\mathcal{H}_M$ and operators acting on fields in the atmosphere. General couplings were considered in [15,16], and linear couplings in [16,17]. Here we explore a model with couplings to the stress tensor $T_{\mu\nu}(x)$,

$$S_{\text{trans}} = \sum_a \int dV \mathcal{A}_a G_a^{\mu\nu}(x) T_{\mu\nu}(x) \ + \text{h.c.} \qquad (2.2)$$

where $dV$ is the near-horizon volume element, and the $G_a$'s are $x$-dependent coefficients. The interaction in general transfers information from the degrees of freedom of $\mathcal{H}_M$ to those of $\mathcal{H}_{\text{near}}$.

Consider, for example, working in an interaction picture where the interaction hamiltonian comes from (2.2) and the remaining evolution is absorbed into that of operators. Then, without (2.2), the state would be of schematic form $|\Psi_0\rangle \approx |0\rangle_U \otimes |\psi\rangle$, where $|0\rangle_U$ describes the Unruh vacuum, and $|\psi\rangle$ the state of matter that formed the BH. Since evolution of the state is frozen in this picture, excitations can be well-described by their "last seen" form near the horizon. With (2.2), the state becomes

$$|\Psi(t)\rangle = T \exp\left\{-i \int dV \left[\mathcal{A}_a(t) G_a^{\mu\nu}(x) T_{\mu\nu}(x) \ + \text{h.c.}\right]\right\} |\Psi_0\rangle \ ; \qquad (2.3)$$

---

[1] In a more natural slicing[23], LQFT may describe some scrambling, but ultimately breaks down.

time dependence of $\mathcal{A}_a$ arises from conjugation with $U(t)$ from (2.1), and from ordinary LQFT evolution.

Whatever the correct detailed description is for internal evolution, the couplings (2.2) need to transfer sufficient information out of the BH. Clearly they would not do so if *e.g.* $\mathcal{A}_a(t)G_a^{\mu\nu}(x) \sim \mathcal{A}_a g^{\mu\nu}$ (with all time dependence in $g_{\mu\nu}$) for all $a$; orthogonal states must map to orthogonal states. If we define operators creating atmosphere excitations

$$T_a^\dagger = \int dV H_a^{\mu\nu}(x) T_{\mu\nu}(x) , \qquad (2.4)$$

for some functions $H_a$, clearly we would like couplings to a sufficiently rich spectrum of the $T_a$'s to encode internal state excitations that we can think of as annihilated by the $\mathcal{A}_a$'s. We can pose this as a condition on the action of the $T_a$'s on the external state $|0\rangle$ of the BH, for example taken to be the Unruh vacuum. We would like

$$\langle 0|T_a|0\rangle = 0 \quad \text{and} \quad \langle 0|T_a T_b^\dagger|0\rangle = \delta_{ab}, \qquad (2.5)$$

so that the $T_a^\dagger$'s create independent excitations, in order to encode sufficient information (the $a=b$ case is a normalization condition).

## 3. Two-dimensional reduction and information encoding

These considerations are most easily investigated in a two-dimensional example. Moreover, for a Schwarzschild BH, this 2d dynamics describes the higher-dimensional dynamics via a reduction in partial waves. Consider for example a scalar field

$$S_\phi = -\frac{1}{2}\int dV \left(g^{\mu\nu}\partial_\mu\phi\partial_\nu\phi + m^2\phi^2\right) \qquad (3.1)$$

in the Schwarzschild background

$$ds^2 = -f(r)dt^2 + \frac{dr^2}{f(r)} + r^2 d\Omega^2 ; \qquad (3.2)$$

in four dimensions, $f = 1 - R/r$, where $R$ is the BH radius. The $D$-dimensional partial wave expansion is

$$\phi = \sum_{Alm}\int \frac{d\omega}{4\pi\omega} b_{\omega lm}^A u_{\omega l}^A(r,t) \frac{Y_{lm}(\Omega)}{r^{D/2-1}} + \text{h.c.} , \qquad (3.3)$$

where the $b_{\omega lm}^A$ are annihilation operators, and $u_{\omega l}^A(r,t)$ are mode solutions in an effective 2d potential (for further details, see *e.g.* [17]). Then, if for example $G^a$ is spherically symmetric, the coupling (2.2) is approximately the same as that to the stress tensors of a collection of 2d fields, labeled by $A, l, m$. (Note this approach fixes a definite time coordinate $t$ at infinity.)

So, we consider an interaction (2.2) in 2d to illustrate basic features of evolution; let us moreover illustrate with the special example of a massless field. Introduce conformal coordinates via tortoise coordinates,

$$ds^2 = f(r)(-dt^2 + dr_*^2) = -f(r)dudv \quad , \quad u = t - r_* \, , \, v = t + r_* \, . \tag{3.4}$$

Then the stress tensor $T_{uu} = \partial_u\phi\partial_u\phi$ for outgoing modes obeys the Virasoro algebra,

$$[T_{uu}, T_{u'u'}] = i(:T_{uu}: + :T_{u'u'}:)\delta'(u-u') - \frac{i}{24\pi}\delta'''(u-u') \tag{3.5}$$

where normal ordering is with respect to positive-frequency modes in $u$ (c.f. (3.3)). This gives us a means to check the conditions (2.5). We will also find it gives a simple way to calculate leading corrections to the energy flux from the BH.

Specifically, let the couplings to the BH internal state be through operators (c.f. (2.4))

$$T_h^\dagger = \int dV h(u,v) T_{uu} = \int \frac{dudv}{2} fh T_{uu} \tag{3.6}$$

for some functions $h(u,v)$; for convenience define $\tilde{h} = fh/2$. For example, let $\tilde{h} = e^{-i\omega u}\delta(v)$, so, for $\omega > 0$,

$$T_\omega = \int du e^{i\omega u} T_{uu} = \int_0^\infty \frac{d\omega'}{4\pi}\left(b_{\omega'}^\dagger b_{\omega+\omega'} - \frac{1}{2}b_{\omega'}b_{\omega-\omega'}\right) \tag{3.7}$$

in terms of creation/annihilation operators like in (3.3), and $T_{-\omega} = T_\omega^\dagger$. Then, for either the Boulware or Unruh vacuum, and $\omega > 0$,

$$\langle 0|T_\omega|0\rangle = 0 \; ; \tag{3.8}$$

for the Boulware vacuum, which satisfies $b_\omega|0\rangle_B = 0$,

$$_B\langle 0|T_\omega T_{\omega'}^\dagger|0\rangle_B = \frac{\omega^3}{12}\delta(\omega - \omega') \, . \tag{3.9}$$

So, the $T_\omega^\dagger$'s produce nontrivial excitations of $|0\rangle_B$. These are eigenstates of the energy

$$:T_0: = \int \frac{d\omega}{4\pi} b_\omega^\dagger b_\omega \tag{3.10}$$

with eigenvalue $\omega$. While $T_\omega|0\rangle_B = 0$ for $\omega > 0$, the positive-frequency $T_\omega$'s do not annihilate $|0\rangle_U$. Excitations of the Unruh vacuum with both higher and lower energies can be created, and generalizations of (3.9) can be calculated with both orderings of $T$, $T^\dagger$.

These excitations thus satisfy analogs of (2.5), and can transfer internal-state information into outgoing modes. Alternately, consider wavepacket $h$'s; for non-overlapping or orthogonal wavepackets these will satisfy similar conditions. In particular, a wavepacket with characteristic $u$-frequency $\omega \sim 1/R$ and $v$-width $R$ has normalization of size (2.5). Creation of excitations via such wavepackets is sufficient to reach the benchmark rate $dS_{\rm vN}/dt \sim -1/R$ necessary for restoration of unitarity[15].

## 4. Energy flux

We wish to understand properties of the state (2.3). An important characteristic is the energy flux. We will approximate this in the *effective source approximation*[16,17], as

$$\langle\Psi(t)|T_{\mu\nu}(t,x^i)|\Psi(t)\rangle \approx \langle\psi(t)|T_{\mu\nu}(t,x^i)|\psi(t)\rangle \tag{4.1}$$

where

$$|\psi(t)\rangle = T\exp\left\{-i\int^t dV' H^{\mu\nu}(x')T_{\mu\nu}(x')\right\}|0\rangle_U \ . \tag{4.2}$$

Here $H^{\mu\nu}$ represents the effective average of $\mathcal{A}_a(t)G_a^{\mu\nu}(x)$ in (2.3) over the pertinent BH internal states: one approximates the effect of the coupling between quantum systems in terms of a classical source acting on the system ($\mathcal{H}_{\rm ext}$) of interest.

To linear order in the source $H$, the flux is

$$\langle\psi(t)|T_{\mu\nu}(x)|\psi(t)\rangle \simeq {}_U\langle 0|T_{\mu\nu}(x)|0\rangle_U - i\int^t dV' H^{\lambda\sigma}(x') \, {}_U\langle 0|[T_{\mu\nu}(x),T_{\lambda\sigma}(x')]|0\rangle_U \ . \tag{4.3}$$

In the 2d example, with

$$|\psi(t)\rangle = T\exp\left\{-i\int^u du'dv'\tilde h T_{u'u'}\right\}|0\rangle_U \ , \tag{4.4}$$

the flux can be calculated in terms of the asymptotic Hawking flux

$$T_{uu}^H(r=\infty) = {}_U\langle 0|{:}T_{uu}{:}(r=\infty)|0\rangle_U = \frac{1}{192\pi R^2} \tag{4.5}$$

via (3.5), giving

$$\langle\psi(t)|T_{uu}(u,v)|\psi(t)\rangle \simeq T_{uu}^H(u,v) + 2T_{uu}^H(\infty)\int^v dv'\partial_u\tilde{h}(u,v') - \frac{1}{24\pi}\int^v dv'\partial_u^3\tilde{h}(u,v') \ . \tag{4.6}$$

For general $h$ the change $\delta_h T_{uu}(u,\infty)$ at infinity will clearly be nonzero and these variations in flux at infinity can carry information. An important question is the change in the *integrated* flux, or radiated energy,

$$\delta_h P_u(u) = \int_{-\infty}^u du'\delta_h T_{uu}(u',\infty) \ . \tag{4.7}$$

Consider, for example, an $h$ that vanishes in the far past and future; then the $h$-dependent contributions to (4.6) integrate to zero. The expression (4.7) shows that while the flux is modulated, at this linear order there is no change in the total radiated energy associated with the escaping information.

One would also like to understand the higher-order flux corrections. The right side of (4.1) is an in-in correlator and can be computed by standard methods (particularly efficient are those of [27]). Such calculations are left for future work, but note the following. Fluctuations about the Unruh vacuum may either raise or lower its energy; *e.g.* consider $T_\omega^\dagger$ or $T_\omega$ acting on $|0\rangle_U$. The former raises the eigenvalue of $:T_0:$ by $\omega$; the latter lowers it by $\omega$. This indicates there can be fluctuations where negative and positive energy contributions to the (quadratic-order) expectation value of $:T_0:$ cancel.

One can calculate the nonlinear change in the flux for a *classical* perturbation $g^{uu}$ of the metric (3.4), via the trace anomaly[28]. The flux is given in terms of the new conformal coordinate

$$u' = u + \frac{1}{4}\int_{-\infty}^v dv' fg^{uu} = u + \int_{-\infty}^v dv'\tilde{h} \ , \tag{4.8}$$

as

$$\langle T_{uu}(u,\infty)\rangle = \left(\frac{\partial u'}{\partial u}\right)^2 T_{uu}^H(\infty) - \frac{1}{24\pi}\left[\frac{\partial_u^3 u'}{\partial_u u'} - \frac{3}{2}\left(\frac{\partial_u^2 u'}{\partial_u u'}\right)^2\right] \ , \tag{4.9}$$

generalizing (4.6). While the nonlinear terms in this expression integrate via (4.7) to be positive definite, the important question is whether interactions (2.2) can achieve small change in radiated flux via the full calculation of the left-hand side of (4.1), taking also into account quantum correlations in the BH state.

While explicit formulas are most easily given for 2d, similar results should hold for the four- or higher-dimensional case, given the spherical reduction to a collection of 2d fields.

These fields move in an effective potential so are no longer massless. However, the primary effect of this is to introduce grey-body transmission and reflection factors. For modes that can efficiently escape the BH, we therefore expect essentially the same considerations to allow comparable transfer of information via couplings of the form (2.2).

## 5. Nonviolence

Another important question is whether a singular horizon[9-12] can be avoided. Section three argued that the necessary information can be transferred by excitations softer than the cutoff, *e.g.* with energies $\sim 1/R$, and with little or no change in the energy flux, suggesting an affirmative answer. However, the change in the state given by (2.3), or approximately (4.2), describes apparent metric fluctuations near the horizon. Let us check that these are innocuous to infalling observers.

We do this in the 2d example, with metric perturbation as in (4.4),

$$ds^2 = -f(r)dudv - \frac{f^2 g^{uu}}{4}dv^2 \ . \tag{5.1}$$

Tidal forces on infalling observers are given, via geodesic deviation, in terms of the Kruskal components of the curvature tensor. These perturbed components may be calculated for (5.1); they are of size comparable to the perturbation in the curvature scalar,

$$\delta \mathcal{R} \sim \partial_u^2 g^{uu} \ . \tag{5.2}$$

For example, if $g^{uu} = h \sim e^{-i\omega u}$, these are of size $\delta \mathcal{R} \sim \omega^2$, or as small as $1/R^2$, which are no larger than curvature of the unperturbed metric. Functions with definite frequency in $u$ are singular, and more regular functions may be more pertinent (*e.g.* analytic functions in Kruskal coordinates), resulting in further suppression in $\delta \mathcal{R}$. These considerations also extend to the higher-dimensional case.

## 6. Conclusion

If information is to escape a BH to unitarize the Hawking process, without modifying the BH thermodynamics as described by the Bekenstein-Hawking entropy $S_{BH}$, apparently the information must escape in fine-grained modulations of radiation preserving average properties of the Hawking flux. We seek a channel for such transfer of information, which is not violent to near-horizon spacetime and infalling observers. This note has investigated information transfer from BH to exterior states via metric perturbation couplings to the stress tensor, (2.2). For a sufficiently rich spectrum of such perturbations, this furnishes the needed channel capacity; a benchmark rate of one qubit emitted per time $R$ is easily achieved. The energy flux can be investigated in a 2d model related to a reduction via spherical harmonics. To linear order in the metric perturbation, a compact support oscillating perturbation produces no extra average energy flux. The higher-order energy flux is yet to be calculated, but is plausibly also small for certain information-carrying couplings; one way to understand this is that certain modulations of the Hawking flux can have unaltered total emitted energy. The needed couplings can be nonviolent to infalling observers. Couplings (2.2) to the stress tensor are also universal, so offer a way to avoid[15,16] the potential problem of producing overfull BHs via mining, since the information-carrying flux increases commensurately with the Hawking flux when a mining channel is opened.[2] While the more fundamental underlying framework may well be based on a structure different from field theory[14], such an effective description may help elucidate some of its essential properties.

### Acknowledgments

I thank R. Bousso for helpful conversations. This work was supported in part by the Department of Energy under Contract DE-FG02-91ER40618 and by Foundational Questions Institute grant number FQXi-RFP3-1330.

---

[2] In fact, the second term in (4.6) is proportional to the outgoing Hawking flux, so where the latter is suppressed, *e.g.* by a small coupling of mining apparatus to atmosphere modes, the former is as well.